\documentclass[aps,prl,twocolumn, longbibliography, superscriptaddress]{revtex4-2}
\usepackage{graphicx,xcolor}
\usepackage{amsmath,amssymb}
\usepackage{physics}
\usepackage{braket}
\usepackage{dsfont}
\usepackage[colorlinks,citecolor=blue,urlcolor=blue]{hyperref}

\usepackage[normalem]{ulem}

\usepackage[many]{tcolorbox}
\newtcolorbox{textbox}{
    boxrule = 0.3pt,
    leftrule = 0.3pt,
    colback = black!0!white,
    colframe = black!75!white,
    breakable
}

\begin{document}

\title{Spontaneous duality breaking and the invertibility of non-invertible symmetries}
\author{José Dupont}
\affiliation{Institute for Theoretical Physics, University of Amsterdam, PO Box 94485, 1090 GL Amsterdam, The Netherlands}
\affiliation{Lorentz Institute for Theoretical Physics, Leiden University, PO Box 9506, 2300 Leiden, The Netherlands}
\author{Jasper van Wezel}
\email{vanwezel@uva.nl}
\affiliation{Institute for Theoretical Physics, University of Amsterdam, PO Box 94485, 1090 GL Amsterdam, The Netherlands}

\date{\today}

\begin{abstract}
Alongside symmetry and topology, duality is used as an organizing principle throughout modern physics. 
Dualities are, however, often not exact and, consequently, implemented by a non-invertible operator, giving rise to a non-invertible symmetry at self-dual quantum critical points. Considering the specific example of the transverse-field Ising model, an archetype for both (self-)duality and non-invertible symmetry, we show that its duality can be made exact by a local adjustment of the boundary conditions. In the modified model with exact duality, the symmetry emerging at the quantum critical point is invertible. Moreover, we find that the exact duality necessitates the presence of an anomalous edge degree of freedom, thus realizing a duality-based rather than topology-based bulk-boundary correspondence. The exactness of the duality furthermore seems to suggest that the spontaneous breaking of a global symmetry can also be described as the spontaneous breakdown of a local dual symmetry. We show that this apparent contradiction of Elitzur’s theorem can be resolved by considering the required symmetry-breaking fields. This moreover leads to the concept of spontaneous duality breaking, in which arbitrarily small interactions with an embedding environment distinguish between mathematically equivalent dual partners, in direct analogy to the spontaneous breakdown of ordinary symmetries. We argue that both duality-based edge modes and spontaneous duality breaking emerge generally across systems in which a local-global duality can be made exact. Together, these phenomena introduce a conceptually novel interplay between duality, symmetry and topology, that is applicable across diverse areas of physics.
\end{abstract}

\maketitle

\section{Introduction}
Together with symmetry and topology, the concept of duality, introduced nearly a century ago by Kramers and Wannier in the context of the 2D classical Ising model~\cite{kramers1941statistics}, has become a cornerstone of modern physics. It underlies our understanding of phenomena ranging from the commonplace correspondence between waves and particles, through proliferation of topological defects in quantum matter, to the relation between different string theories, and even the equivalence of gravitational and conformal field theories~\cite{kosterlitz1973ordering, Fradkin_2017,savit1980duality, witten1995string, maldacena1999large, aasen2020topologicaldefectslatticedualities, choi2022noninvertible, seibergsymmetries}. Moreover, local-global dualities, with the transverse-field Ising model as an archetypal example, allow the study of otherwise unsolvable strongly-coupled theories using perturbation theory on their dual versions~\cite{fradkin1978order, seiberg1994electric, fisher1990quantum, Beekman2017}.

\begin{figure}[tb!]
    \centering
    \includegraphics[width=0.8\columnwidth]{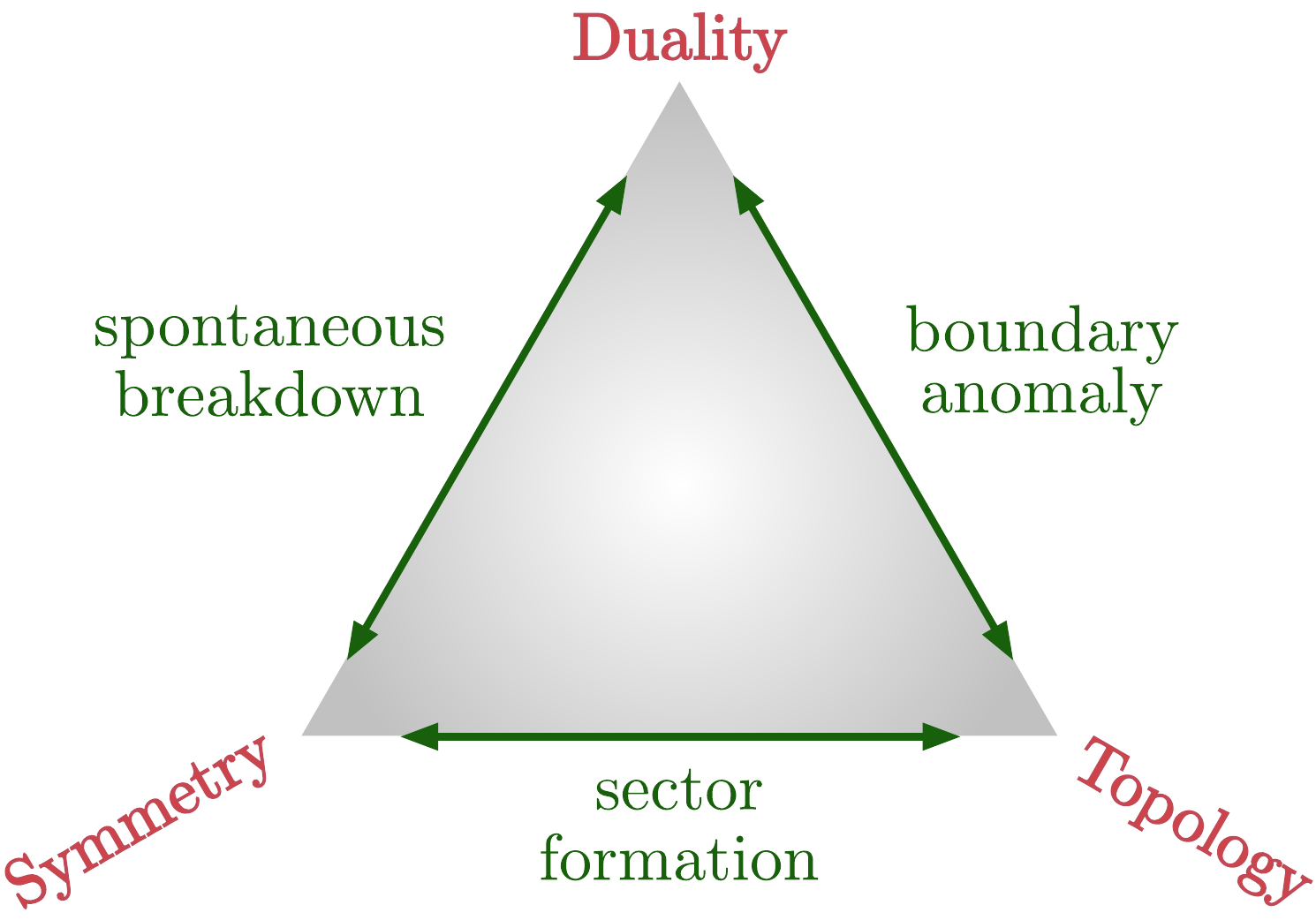}
    \caption{{\bf Duality, symmetry and topology.} Symmetry and topology both constrain dynamics to sectors in which either the order parameter or topological charge is conserved. Similarly, topology and duality, once made exact, both contain a type of bulk-boundary correspondence necessitating the emergence of anomalous edge degrees of freedom. Exact duality and invertible symmetry, finally, coincide in the diverging susceptibility to environmental perturbations that they give rise to, which causes both to tend towards a type of spontaneous breakdown. All three constrain the phases of matter realized in nature, by distinguishing between stable and unstable states.}
    \label{fig:triangle}
\end{figure}

In the triad of organizing principles, duality traditionally stands somewhat apart from symmetry and topology. Symmetry and topology are both associated with a type of rigidity, constraining dynamics to sectors in which either an order parameter orientation or topological indices are conserved (see Fig.~\ref{fig:triangle}). Similarly, both are associated with emergent degrees of freedom, either at spatial boundaries or in the long wavelength limit~\cite{ssb_book}. Duality on the other hand, at first sight seems a purely mathematical construction. It may unveil an alternative description or interpretation of a given phenomenon, elucidate the emergence of symmetry at a quantum critical point, or be implemented in terms of topological defect operators, but it does not seem to yield physical consequences like associated degrees of freedom, or phases of its own. Here, we argue that this apparent limitation is due to the fact that dualities as commonly encountered, are often not exact. We show by example of the transverse-field Ising model that duality, once made exact, gives rise to anomalous boundary degrees of freedom that are directly analogous to those found in topologically nontrivial materials, and that duality can be spontaneously broken by infinitesimal environmental interactions just like symmetries. As shown schematically in Fig.~\ref{fig:triangle}, these conceptual overlaps complete the triangle of duality, topology, and symmetry, and put them on equal footing.

\section{The transverse-field Ising model}
Although we argue the emergence of duality-based boundary degrees of freedom and the spontaneous breakdown of duality to be universal features of the broad class of local-global dualities~\cite{Beekman2012,BeekmanPhD,fradkin1978order, seiberg1994electric, fisher1990quantum, Beekman2017}, we introduce them here in the particular setting of the transverse-field Ising model. This quantum model inherits its self-duality from the 2D classical Ising model via the transfer-matrix method~\cite{schultz1964two, kogut1979introduction}, which in turn is the original model used by Kramers and Wannier to introduce the concept of duality~\cite{kramers1941statistics}. They famously showcased how dualities can be used as a powerful tool for exploring phase diagrams, leveraging the duality to identify the precise critical point of the 2D Ising model without exact solutions or numerical methods~\cite{kramers1941statistics, Fradkin_2017}.

More recently, dualities have been used to identify so-called non-invertible symmetries~\cite{frohlich2004kramers, choi2022noninvertible, mcgreevy2023generalized}. 
These occur because an operator implementing a self-duality leaves the Hamiltonian invariant at the critical point, and thus embodies a symmetry. However, because most dualities, including the Kramers-Wannier duality, are in fact approximate, the operators implementing them are non-invertible~\cite{moore1989naturality}. In particular, this construction has been shown to apply to the transverse-field Ising model, which has a non-invertible symmetry at its critical point~\cite{aasen_fendley_2016, seibergsymmetries}. Here, we demonstrate that the duality of the transverse-field Ising model at finite sizes is approximate because of boundary effects, but can be made exact by introducing an anomaly at the boundary. This exact duality produces an invertible symmetry at the critical point which persists to the thermodynamic limit and gives rise to duality-based boundary states as well as a diverging susceptibility to environmental perturbations.

\section{Approximate Self-Duality}
The Hamiltonian for the transverse-field Ising model with transverse field strength $\lambda$ is given by~\cite{schultz1964two, pfeuty1970one, kogut1979introduction}:
\begin{equation}\label{eq.ising_model_transverse_field}
    \hat{H}(\hat{\sigma};\lambda) = - \sum_{n}\left( \hat{\sigma}^z_n\hat{\sigma}^z_{n+1} + \lambda \hat{\sigma}^x_n \right).
\end{equation}
Here $\hat{\sigma}_n^{x,z}$ denotes a Pauli $x$ or $z$ spin operator on lattice site $n \in \{1,\dots,N\}$, and the ferromagnetic coupling strength is set to one. The Hamiltonian is invariant under a simultaneous flip of all spins, implemented by the operator $\hat{X}_{\sigma}^{\phantom{x}} = \prod_{n} \hat{\sigma}_n^x$.
If the transverse field $\lambda$ is weak compared to the ferromagnetic coupling, the Hamiltonian has a ferromagnetic ground state that spontaneously breaks the global spin-flip symmetry.
Conversely, if the transverse field is strong compared to the ferromagnetic coupling, the model has a unique paramagnetic ground state in which the total magnetization vanishes~\cite{pfeuty1970one}.

An alternative description of the model can be given in terms of dual operators defined as~\cite{kogut1979introduction}:
\begin{align}
    \hat{\mu}_n^x & = \hat{\sigma}_n^z \hat{\sigma}_{n+1}^z, &
    \hat{\mu}_n^z & = \prod_{m \leq n} \hat{\sigma}_m^x. \label{eq.ising_original_to_dual3}
\end{align}
Here, the boundary conditions $\hat{\sigma}^{x,z}_{N+1} = e^{i\theta}\hat{\sigma}^{x,z}_1$ with either $\theta=0$ (periodic) or $\theta \neq 0$ (partially twisted) are used to define $\hat{\mu}_n^x$ for all values of $n$. 
While being nonlocal operations in terms of $\hat{\sigma}$-operators, from the dual perspective, the operators $\hat{\mu}_n^{x,z}$ are ordinary Pauli spin operators, representing the $x$ and $z$ components of a dual spin on site $n$. Indeed, for bulk values $1 < n < N$, the operators $\hat{\mu}_n^{x,z}$ square to the identity and satisfy commutation relations expected of Pauli spin operators~\cite{kogut1979introduction}. Using Eq.~\eqref{eq.ising_original_to_dual3}, we can rewrite the Ising Hamiltonian in terms of $\hat{\mu}$-spins, and find:
\begin{align}\label{eq.self-dual_H_approximate}
    \hat{H}(\hat{\sigma};\lambda)
     = \lambda \hat{H}(\hat{\mu};1/ \lambda).
\end{align}
Because of this relation, the model is considered to be self-dual: expressing the transverse-field Ising Hamiltonian in terms of dual spin operators results in another version of the same transverse-field Ising model but with inverse field-strength.

Importantly, $\hat{H}(\hat{\mu})$ is of precisely the same form as the original model only if the $\hat{\mu}$-operators behave as Pauli spin matrices. This, however, is not the case for the $\hat{\mu}$-operators at the boundaries of the 1D chain. To wit, $[\hat{\mu}_N^x,\hat{\mu}_1^z] \neq 0$ even though these operators act on different lattice sites. Moreover, the $\hat{\mu}$-operators do not satisfy the boundary conditions imposed on the $\hat{\sigma}$-operators~\cite{Fradkin_2017}, and the global $\hat{\mu}$-spin flip operator $\hat{X}_{\hat{\mu}}= \prod_n \hat{\mu}_n^x$ is proportional to the identity, resulting in different ground state degeneracies of the original and dual Hamiltonians in the thermodynamic limit. Despite the fact that the Hamiltonians $\hat{H}(\hat{\sigma};\lambda)$ and $\hat{H}(\hat{\mu};\lambda)$ look the same up to replacement of $\hat{\sigma}$-operators with $\hat{\mu}$-operators, they thus have different spectra~\cite{aasen_fendley_2016} (see the Appendix for details).

The original argument by Kramers and Wannier posited that the spectrum $E_{\hat{\sigma}}(\lambda)$ of $\hat{H}(\hat{\sigma};\lambda)$ equals the spectrum $E_{\hat{\mu}}(\lambda)$ of $\hat{H}(\hat{\mu};\lambda)$. In that case, Eq.~\eqref{eq.self-dual_H_approximate} would imply $E_{\hat{\sigma}}(\lambda) = \lambda E_{\hat{\mu}}(1/\lambda) = \lambda E_{\hat{\sigma}}(1/\lambda)$. The mass gap would then follow a similar relation resulting in a critical transverse field strength of $\lambda_c=1$~\cite{kramers1941statistics,schultz1964two,kogut1979introduction}. Because the original and dual operators do not satisfy the same algebraic relations, however, the corresponding spectra cannot be identified~\cite{Fradkin_2017}. The reason that the original conclusion $\lambda_c=1$ still holds, is that in the thermodynamic limit the deviating behavior of the $\hat{\mu}$-operators at the boundary becomes irrelevant. Local perturbations to the transverse-field Ising Hamiltonian affect the order parameter only locally, and do not change its bulk properties~\cite{wezel_ssb_introduction,ssb_book}. The correct critical behavior of the model can thus be derived based on an approximate duality~\cite{Fradkin_2017}.

In stark contrast to its negligible influence on the mass gap, the fact that the duality is approximate does significantly impact its implementation in terms of a duality transformation. An operator $\hat{D}$ defined such that $\hat{D} \hat{H}(\hat{\sigma};\lambda) = \hat{H}(\hat{\mu};\lambda) \hat{D}$, satisfies the relations~\cite{Seiberg_2024, aasen_fendley_2016}:
\begin{align}\label{eq.D_action_on_sigma}
        \hat{D} \hat{\sigma}_n^x &= \hat{\sigma}_n^z\hat{\sigma}_{n+1}^z \hat{D} = \hat{\mu}_n^x \hat{D} \notag \\
        \hat{D} \hat{\sigma}_n^z\hat{\sigma}_{n+1}^z &= \hat{\sigma}_{n+1}^x  \hat{D} = \hat{\mu}_n^z\hat{\mu}_{n+1}^z \hat{D},
\end{align}
As detailed in the Appendix, the approximate nature of the duality and the consequent difference between original and dual spin-flip operators necessarily renders $\hat{D}$ a projection operator, and hence non-invertible.

If the duality had been exact, the original and dual Hamiltonian at the critical value $\lambda=1$ would have been the same, and the operator $\hat{D}$ would have commuted with the Hamiltonian at that point. Because of this, $\hat{D}$ is commonly referred to as a non-invertible symmetry of the model at criticality~\cite{seibergsymmetries}.

\section{Exact Self-Duality}
The non-invertible operator $\hat{D}$ is often treated in a manner comparable to usual, invertible symmetries, by introducing so-called superselection sectors~\cite{aasen_fendley_2016, Seiberg_2024, shao2024whatsundonetasilectures}. For the transverse-field Ising model with boundary condition $\hat{\sigma}_{N+1}^{x,z} = e^{i\theta}\hat{\sigma}_1^{x,z}$, such a sector consist of all states satisfying $\hat{X}_{\sigma}\ket{\psi}=e^{i\theta}\ket{\psi}$. Within this restricted set of states, $\hat{D}$ is invertible (see the Appendix). The argument that it is not possible to evolve from one superselection sector to another without breaking the global spin-flip symmetry is then commonly evoked to restrict attention to just a single sector~\cite{aasen_fendley_2016, Seiberg_2024, shao2024whatsundonetasilectures, Lootens_2023, kadanoff1971determination}.

Here, we take a different perspective. Any physical implementation of the model will necessarily be embedded in an environment, and symmetry-breaking perturbations and resulting superpositions over superselection sectors may naturally occur. For example, the all-up state is a superposition of states in two distinct sectors. Moreover, such states may be purposely created. Because for any given choice of $\theta$, the duality operator is invertible only on a single superselection sector, the unavoidable presence of multiple sectors in implementations of the model prevents any interpretation of $\hat{D}$ as a physical operator, even in the thermodynamic limit. Instead, we will therefore construct an exact duality transformation by focusing on open boundary conditions. This exact duality applies to any system size, to all possible states of the system, and is implemented by an invertible operator.

To construct the exact duality, we first require a set of dual operators $\hat{\rho}_n^{x,z}$ that are exactly, rather than approximately, Pauli-spin operators. This is possible by redefining Eq.~\eqref{eq.ising_original_to_dual3} for only a single operator:
\begin{equation}
    \hat{\rho}_n^x = \begin{cases}
        \hat{\sigma}_n^z\hat{\sigma}_{n+1}^z, & n \neq N
        \\ 
        \hat{\sigma}_n^z & n = N
    \end{cases} \qquad
    \hat{\rho}_n^z = \prod_{m \leq n} \hat{\sigma}_m^x. \label{eq.ising_original_to_dual3_open}
\end{equation}
This definition is for $n\in\{1\dots N\}$ and does not require any boundary condition. Because these dual operators obey the same algebraic relations as the original $\hat{\sigma}$-operators, there is a unitary, invertible operator $\hat{U}$ such that $\hat{\rho}_n^{x,z} = \hat{U}^{\dagger} \hat{\sigma}_n^{x,z} \hat{U}$. This coincides with one of the duality operators found using a bond-algebraic approach to dualities in Ref.~\cite{ortiz2026generalized}. The operator $\hat{U}$ is defined in the Appendix.

Writing the transverse-field Ising model with open boundary conditions in terms of $\hat{\rho}$-operators, it is found not to be self-dual. The self-duality can be restored, however, by either adding a local symmetry-breaking term $\hat{\sigma}_N^z$, or by removing the operator $\hat{\sigma}_1^x$:
\begin{align*}
    \hat{H}'(\hat{\sigma};\lambda) &=
    -\sum_{n = 1}^{N-1} \hat{\sigma}_n^z\hat{\sigma}_{n+1}^z - \uline{\hat{\sigma}_N^z}
    -\lambda \left(\uwave{\hat{\sigma}_1^x}+ \sum_{n = 2}^N \hat{\sigma}_n^x \right)  \\
    &=
    -\lambda \left(\uwave{\hat{\rho}_1^z} + \sum_{n = 1}^{N-1} \hat{\rho}_n^z\hat{\rho}_{n+1}^z 
    + \frac{1}{\lambda}\left(\sum_{n = 1}^{N-1} \hat{\rho}_n^x + \uline{\hat{\rho}_N^x} \right)
    \right)
    \\
    &= \lambda \hat{H}'(\hat{\rho};1/\lambda).
\end{align*}
The final line is obtained by relabeling the lattice sites $n \leftrightarrow N-n$. The underlines and wiggles indicate which terms in the Hamiltonian are related by Eq. \eqref{eq.ising_original_to_dual3_open}. The self-duality is exact if the underlined and wiggled term are either both included or both excluded, where the former choice coincides with the Hamiltonian derived through a bond-algebraic approach to dualities~\cite{cobanera2011bond}. Including them, the transverse field acts on all dual lattice sites, but the global $\hat{X}_{\hat{\sigma}}$ symmetry is broken. The resulting Hamiltonian has a unique, non-degenerate ground state for all values of $\lambda$ and $N$, and no symmetry-breaking phase transition. We therefore restrict attention to the more interesting case of excluding the underlined and wiggled terms.

\begin{figure}[tb]
    \centering
    \includegraphics[width=\columnwidth]{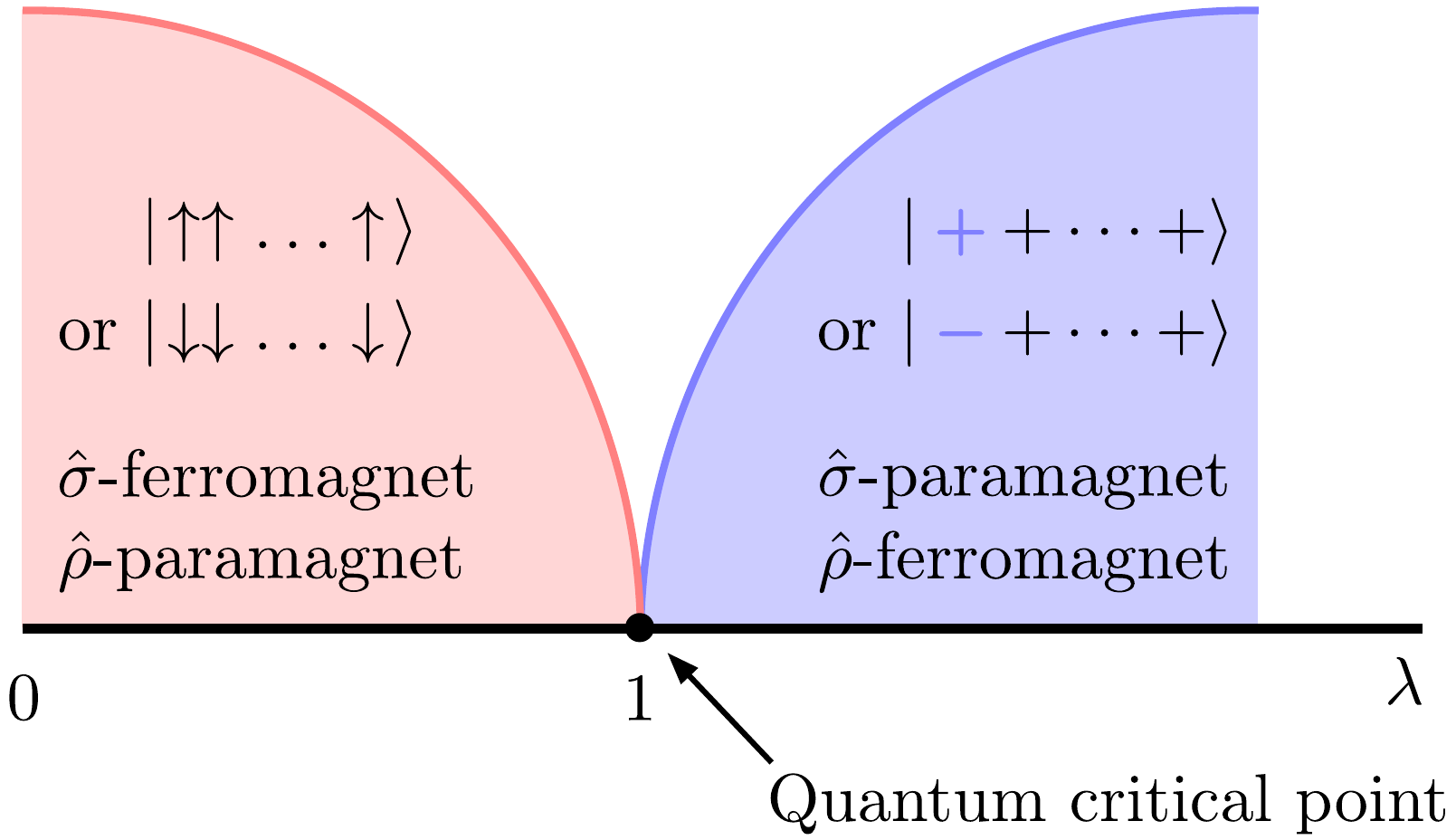}
    \caption{{\bf Schematic phase diagram of the transverse-field Ising model with anomaly.} The states indicated are adiabatically connected to the ground states in the thermodynamic limit throughout the corresponding phases and become exact ground states as either $\lambda$ or $1/\lambda$ vanishes.
    Unlike the model without anomaly, the dual Hamiltonian has two ground states in the thermodynamic limit for $\lambda>1$. These are related by the global symmetry $\hat{X}_{\hat{\rho}}$, which corresponds to a local symmetry $\hat{\sigma}_1^z$ acting on the anomalous boundary degree of freedom.}
    \label{fig:phase_diagram_exact}
\end{figure}

The exactly self-dual Hamiltonian we thus consider is:
\begin{equation}\label{eq.H_with_symmetry}
    \hat{H}'(\hat{\sigma};\lambda) = -\sum_{n = 1}^{N-1} \hat{\sigma}_n^z\hat{\sigma}_{n+1}^z - \lambda \sum_{n = 2}^N \hat{\sigma}_n^x.
\end{equation}
This Hamiltonian retains the $\hat{X}_{\hat{\sigma}}$ symmetry, and is also invariant under the global $\hat{\rho}$-spin-flip symmetry $\hat{X}_{\hat{\rho}} = \prod_{n} \hat{\rho}_n^x$. The presence of a global symmetry in $H'(\hat{\rho};\lambda)$ allows the dual model to have a degenerate ground state susceptible to spontaneous symmetry breaking in the thermodynamic limit, in direct correspondence with the original model. As shown schematically in Fig.~\ref{fig:phase_diagram_exact}, the all-$\hat{\rho}$-spin-up and all-$\hat{\rho}$-spin-down states written in terms of $\hat{\sigma}$-spins correspond to having either all of them in the symmetric configuration $\ket{+}=\tfrac{1}{\sqrt{2}}(\ket{\uparrow}+\ket{\downarrow})$, or the same state with only the spin on site $n=1$ converted to the antisymmetric state $\ket{-}=\tfrac{1}{\sqrt{2}}(\ket{\uparrow}-\ket{\downarrow})$. 

In terms of the original $\hat{\sigma}$-operators, the global symmetry $\hat{X}_{\hat{\rho}}$ corresponds to a local symmetry $\hat{\sigma}_1^z$. Insisting on the exactness of the bulk duality and the global symmetry thus enforces the presence of a free local degree of freedom at the boundary site $n=1$, where the transverse field in Eq.~\eqref{eq.H_with_symmetry} does not act. We refer to this degree of freedom as a duality anomaly. It is analogous to topological anomalies in for example Chern-Simons theory when insisting on the exactness of bulk gauge invariance in the presence of open boundary conditions~\cite{witten1989quantum, wen1992theory}. 

\section{Conformal Behavior}
The 1D transverse-field Ising model and the 2D classical Ising model give rise to the same conformal behavior at their critical points, which is characterized by the fusion rules~\cite{aasen_fendley_2016}:
\begin{equation}\label{eq.cft}
    \eta\otimes\eta = \mathds{1}, \qquad  D \otimes \eta = D, \qquad D\otimes D = \mathds{1}+\eta.
\end{equation}
Here, $\eta$, $D$, and $\mathds{1}$ are primary operators in the conformal field theory (CFT) describing the model at criticality. Each can be identified with a topological operator in the 1D chain~\cite{aasen_fendley_2016, Seiberg_2024}. The operator $\eta$ is a so-called spin-flip defect, and corresponds to the operator that flips all spins to its left; $D$ is called the duality defect, and corresponds directly to the operator $\hat{D}$ defined by Eq.~\eqref{eq.D_action_on_sigma}; $\mathds{1}$, finally, is the identity operator.

The duality defect satisfies the fusion rule $D\otimes D = \mathds{1}+\eta$, which implies that it is a non-invertible operator in the CFT~\cite{aasen_fendley_2016}. This is consistent with the CFT arising as the continuum limit of a critical model with periodic boundary conditions, for which we have seen that the duality is not implementable by an invertible symmetry. Since the fusion algebra is intrinsic to the bulk critical theory~\cite{frohlich2007duality}, introducing open boundaries in the CFT can affect how a defect acts on a boundary, but does not alter its bulk fusion rules, and does not turn the bulk duality defect $D$ into an invertible operator. Rather, the operator $\hat{U}$ is invertible precisely because it acts on the full microscopic system, including boundary information that is absent from the bulk CFT description. 

The models with and without exact duality must share an effective description for their bulk physics in the thermodynamic limit, because they differ only locally. They therefore fall in the same universality class, described by the same CFT and the same (non-invertible) fusion rules. An intuitive way to interpret this, is to note that the mass gap of both models is determined by the energy difference between states that differ by a topological bulk excitation (a kink in the 1D chain), which cannot vanish due to only local perturbations.

\section{Spontaneous Duality Breaking}\label{sec.ssb_local_symmetry}
In the modified transverse-field Ising model with anomaly, the global symmetry $\hat{X}_{\hat{\sigma}}$ is spontaneously broken in the phase with $\lambda<1$. As usual, the terminology `spontaneous' indicates that the sensitivity of the ground state to small symmetry-breaking perturbations in the Hamiltonian diverges in the thermodynamic limit~\cite{koma1994symmetry,wezel_ssb_introduction, ssb_book}. The global $\hat{X}_{\hat{\sigma}}$ symmetry, however, can equivalently be described as a local $\hat{\rho}_N^z$ symmetry. From the dual perspective, it thus seems like a local symmetry is spontaneously broken in the $\lambda<1$ phase. This is surprising, as it seems to contradict Elitzur's theorem, which states that local symmetries can never be spontaneously broken~\cite{elitzur1975impossibility,wezel_ssb_introduction, ssb_book}. For the phase with $\lambda>1$, the same situation occurs in reverse, as we expect a paramagnetic phase of $\hat{\sigma}$-spins with an unbroken local symmetry, but we find a symmetry-breaking ferromagnetic phase in terms of $\hat{\rho}$-spins.

The resolution to this paradox lies in what is meant by spontaneously breaking a symmetry. A local perturbation of the form $-\epsilon \hat{\sigma}^z_1$ suffices to render one particular $\hat{\sigma}$-ferromagnetic state the unique ground state in the $\lambda<1$ phase, for any value of $\epsilon$. Similarly, the perturbation $-\epsilon \hat{\rho}^z_1 = -\epsilon \hat{\sigma}^x_1$ suffices to pick out one particular state breaking the local $\hat{\sigma}$-symmetry in the $\lambda>1$ phase. Both the local and global symmetry can be broken by perturbations with arbitrarily small values of $\epsilon$. The resulting symmetry-breaking states, however, differ in terms of their stability.

Of the equivalent ways in which stability of states can be defined~\cite{ssb_book}, stability against local measurement is most convenient for the present discussion. Roughly speaking, an ordered state is then considered stable if for sufficiently large systems the value of the order parameter expectation value cannot be influenced significantly by the measurement of any local observable~\cite{ssb_book,stability}. This is the case for a ferromagnetic configuration of $\hat{\sigma}$-spins in the transverse-field Ising model, for which local measurement of a single spin in any basis has a negligible effect on the total magnetization of the entire system in the limit of large system size. The broken local symmetry $\hat{\sigma}_1^z$ in the paramagnetic phase on the other hand is not stable against local measurement, because the distinction between its two symmetry-breaking configurations consists of only a single $\hat{\sigma}$-spin-flip at site $n=1$. In practice this implies that even if the local symmetry is broken at a given instant of time, it can be immediately restored by local fluctuations from an external environment~\cite{ssb_book,stability}.

Of course, the same discussion of stability against local measurement or fluctuations can be had in terms of $\hat{\rho}$-operators. It thus seems that there is always a local measurement, in terms of either $\hat{\sigma}$ or $\hat{\rho}$-operators, that significantly affects any ordered state. The apparent conclusion would be that no stable long-range order is possible in the transverse-field Ising model at any value of $\lambda$. This reasoning, however, ignores a crucial aspect inherent to any mathematical description of a physical system: to make predictions about observables in the real world, the mathematical operators appearing in the model need to be assigned a real-world implementation. That is, if the transverse-field Ising model is to describe a 1D magnetic material, then either the $\hat{\sigma}$ or the $\hat{\rho}$-operators represent physical spin degree of freedoms that are local in the experimental setup used. The operators dual to them obey the algebra of spin degrees of freedom, but their implementation in the lab is necessarily non-local. The same holds for the local measurements or fluctuations considered above. They may look local in terms of either $\hat{\sigma}$ or $\hat{\rho}$-operators, but in any physical implementation of the mathematical model, only one type of measurement or fluctuation is actually local in the laboratory setting.

The exact duality of the modified transverse-field Ising model brings to the fore the crucial role of the physical embedding. In any laboratory setting, the physical implementation of the transverse-field Ising model (with anomaly) will necessarily involve weak interactions with the environment, which define locality of operators and thereby distinguish between the dual descriptions of the model.

In this way, the duality of the mathematical model is broken by perturbations that may be immeasurably weak and practically uncontrollable, but that nonetheless have a qualitative effect on the observed bulk configuration of the physically implemented system. Because of the close analogy to how immeasurable and uncontrollable environmental influences cause symmetries to break, we refer to the distinction between dual observables being made by its physical realization as spontaneous duality breaking.

\section*{Discussion}
We showed that the self-duality of the transverse-field Ising model with (twisted) periodic boundary conditions is approximate in the sense that the dual operators do not obey the same algebra as the original spins, the original and dual Hamiltonian have distinct symmetries, and their spectra and number of ground states in the thermodynamic limit differ. Considering open boundary conditions, we showed that it is possible to impose an exact duality on the transverse-field Ising model. Because it is exact, the duality can be implemented by an invertible operator, and the symmetry arising at the critical point is a conventional, invertible symmetry. 

Moreover, we found that enforcing the bulk duality to be exact necessitates the presence of an anomalous degree of freedom at the boundary. The resulting bulk-boundary correspondence and localized anomaly are analogous to those found in topological insulators, but arise here from duality instead of topology. We stress that the conformal continuum description of the model is not affected by the local modifications at the boundary, which have vanishing influence in the thermodynamic limit.

We finally pointed out that the modified model with an exact duality is capable of spontaneously breaking a symmetry on both sides of its critical point. This is surprising at first sight, since on both sides the broken symmetry is local in terms of either the original or the dual operators. This paradox is resolved by embedding the system in an environment. The subsequent interactions, no matter how weak, cause fluctuations that are local in the laboratory. In this way, they distinguish between the dual descriptions and provide the fluctuations that prevent a state with broken local symmetry from being stable, even if the symmetry is global in terms of dual operators.
In analogy to the way symmetries may be broken by vanishingly weak environmental influences, we dub the distinction between mathematically equivalent dual descriptions due to its physical embedding, spontaneous duality breaking.

Although the discussion so far focused on the interplay between duality, symmetry, and topology in the specific context of the transverse-field Ising model, we argue that the results are universal throughout the broad class of local-global dualities~\cite{fradkin1978order, seiberg1994electric, fisher1990quantum, Beekman2017,Beekman2012,BeekmanPhD}. This type of duality maps a theory with global symmetry to one with local degeneracies, which will act as boundary degrees of freedom if the duality is made exact. These introduce a local symmetry which will be global in the dual description, but which will not be spontaneously broken. This constitutes spontaneous duality breaking.

The impact of spontaneous duality breaking and the emergence of edge anomalies as a consequence of exact duality may be expected to extend beyond simple toy models like the transverse-field Ising magnet.
For example, a famously and explicitly broken duality in nature is that between electric and magnetic fields. The existence of magnetic monopoles would restore this duality and make it exact~\cite{dirac1931quantised}. Interpreting duality as a property of idealized models that can be broken by external influences, as in the case of spontaneous duality breaking described above, suggests the possibility of the electromagnetic duality being broken by an as-yet unexplored physical field inhibiting the creation of magnetic monopoles.
Another example with more direct practical relevance, may lie in utilizing the boundary degrees of freedom emerging due to exact duality to create decoherence-free subspaces in quantum computation setups. If one implements an exactly dual model from qubits that are local in the lab, manipulating and controlling a single qubit is equivalent to steering a large, nonlocal superposition in the dual degrees of freedom. By the same token, any computation on the dual degrees of freedom is subjected to decoherence only through a single physical qubit, which may be arbitrarily well isolated. This is similar to, but distinct from the isolation employed in setups such as the toric code~\cite{kitaev2003fault}. Rather than being based in delocalized topological degrees of freedom, the protection of quantum information in this case is rooted in the necessarily free boundary degrees of freedom in realizations of spontaneously broken duality.

\section{Acknowledgments}
The work of J.D. was supported by the Netherlands Organization for Scientific Research (NWO/OCW), as part of Quantum Limits (project number SUMMIT.1.1016).

\section{Appendix}

\subsection{A: Inexact duality with periodic boundaries}
Using periodic boundary conditions to define the model Hamiltonian $\hat{H}(\hat{\sigma})$, its dual version $\hat{H}(\hat{\mu})$ does not exactly represent a transverse-field Ising model with periodic boundary conditions in its own right, because the $\hat{\mu}$-operators at the boundaries $n=1$ and $n=N$ do not behave like Pauli operators. Even though they act on different lattice sites, the operators $\hat{\mu}_N^x$ and $\hat{\mu}_1^z$ do not commute:
\begin{equation}\label{eq.mu_nonzero}
    [\hat{\mu}_N^x, \hat{\mu}_1^z]= e^{i\theta} \hat{\sigma}_N^z [\hat{\sigma}_1^z,\hat{\sigma}_1^x]  \neq 0.
\end{equation}
Moreover, the boundary conditions imposed on the $\hat{\sigma}$-operators are not satisfied by the $\hat{\mu}$-operators:
\begin{align}\label{eq.no_periodic_bc}
    \hat{\mu}_{N+1}^z = \prod_{m\leq N+1} \hat{\sigma}_m^x = e^{i\theta} \prod_{m=2}^N \hat{\sigma}_m^x \neq e^{i \theta} \hat{\sigma}_1^x = e^{i \theta} \hat{\mu}_1^z.
\end{align}
Finally, the global $\hat{\mu}$-spin-flip operator $\hat{X}_{\hat{\mu}}$ satisfies:
\begin{align}\label{eq.m_1=1}
    \hat{X}_{\hat{\mu}} &= \prod_{n=1}^N \hat{\mu}_n^x 
    = \hat{\sigma}_1^z\hat{\sigma}_{N+1}^z
    = e^{i\theta} \hat{\mathds{1}}.
\end{align}
Since $\hat{X}_{\hat{\mu}}$ is proportional to the identity operator, it represents a trivial transformation on the $\hat{\mu}$-spin configuration. It is thus not a physical symmetry operator, even though it commutes with the Hamiltonian. As a consequence, the dual Hamiltonian $\hat{H}(\hat{\mu})$ with periodic boundary conditions has no global spin-flip symmetry, in contrast to the original model. This implies that the two Hamiltonians have different spectra. For $\lambda < 1$, the Hamiltonian $\hat{H}(\hat{\sigma};\lambda)$ has two degenerate ground states in the thermodynamic limit, which connect to the all-spin-up and all-spin-down configurations in the $\lambda\to 0$ limit. The same Hamiltonian expressed in terms of $\hat{\mu}$-spins is $\hat{H}(\hat{\mu};\lambda) = \lambda \hat{H}(\hat{\sigma};1/ \lambda)$. This Hamiltonian has a unique ground state that connects to the state with all $\hat{\sigma}$-spins in the symmetric configuration $\tfrac{1}{\sqrt{2}}(\ket{\uparrow}+\ket{\downarrow})$ in the $\lambda \to 0$ limit. 

\subsection{B: Non-invertible duality operator}
To show that the operator $\hat{D}$ implementing the duality is a projector, consider a state $\ket{\psi}$ that behaves as $\hat{X}_{\hat{\sigma}}\ket{\psi} = e^{i\varphi}\ket{\psi}$ under the global $\hat{\sigma}$-spin-flip. With boundary conditions $\hat{\sigma}_{N+1}^{x,z}=e^{i\theta}\hat{\sigma}_1^{x,z}$, using first Eq.~\eqref{eq.D_action_on_sigma} and then Eq.~~\eqref{eq.m_1=1} yields:
\begin{align}\label{eq:noninvproof}
    e^{i\varphi}\hat{D} \ket{\psi} & =  \hat{D} \hat{X}_{\hat{\sigma}} \ket{\psi} = \hat{X}_{\hat{\mu}} \hat{D} \ket{\psi} = e^{i \theta} \hat{D} \ket{\psi}.
\end{align}
For any state with $\theta\neq\varphi$, this equation implies $\hat{D} \ket{\psi}=0$, so that $\hat{D}$ has non-zero kernel and thus cannot be invertible.

However, $\hat{D}$ is invertible on the subset of states for which $\varphi=\theta$. This defines a superselection sector given a value of $\theta$ set by the boundary conditions.

\subsection{C: Invertible duality operator}
Using open boundary conditions, the duality can be implemented by an invertible operator.  Contrasting Eq.~\eqref{eq.mu_nonzero} for the $\hat{\mu}$-operators defined with periodic boundaries, in this case we require $[\hat{\rho}^x_N,\hat{\hat{\rho}}_1^z]=0$ of the dual $\hat{\rho}$-operators. The requirement is fulfilled by taking $\hat{\rho}_N^x= e^{i\varphi} \hat{\sigma}_N^z$, where Hermiticity implies $\varphi = 0, \pi$ and we choose $\varphi=0$ without loss of generality.

Once the duality is made exact by local redefinitions of the dual operators in Eq.~\eqref{eq.ising_original_to_dual3_open} and of the Hamiltonian in \eqref{eq.H_with_symmetry}, it can be implemented by an invertible operator $\hat{U}_N$ for a transverse-field Ising model with $N$ sites. The operator $\hat{U}_N$ can be constructed inductively via the recurrence relation:
\begin{align}
    \hat{U}_N  &=(\hat{\mathds{1}}_1 \otimes\hat{U}_{N-1}) \cdot \hat{T}_{1,2}, \\
    \hat{T}_{1,2} &= \frac{1}{2\sqrt{2}}  \left[ (\hat{\mathds{1}}_1 + \hat{\sigma}_1^x)(\hat{\sigma}_2^x + \hat{\sigma}_2^z) \right. \notag \\
    & ~~~~~~~~~~~~~~~~ \left. +
    (\hat{\mathds{1}}_1 - \hat{\sigma}_1^x) (\hat{\mathds{1}}_2 +\hat{\sigma}_2^z\hat{\sigma}_2^x) \right].  \notag 
    \label{eq:recursive}
\end{align}
Here we defined the initial transformation on the $N=2$ model, written in the eigenbasis of the $\hat{\sigma}^z$-operators, as:
\begin{align}
        \hat{U}_2  & = \frac{1}{2}\begin{pmatrix}
        1&1&1&1 \\ 1&-1&1&-1\\1&-1&-1&1\\1&1&-1&-1
    \end{pmatrix}.
\end{align}
Noticeably, it is also possible to find a set of exactly dual operators using the bond-algebraic approach to dualities~\cite{cobanera2010unified, cobanera2011bond}, which led Ref.~\cite{ortiz2026generalized} to the explicit form, which is equivalent to Eq.~\eqref{eq:recursive}:
\begin{equation}
    \hat{U}_N=\left(\prod_{j=1}^{N-1} \mathrm{C}_{j+1, j}^x\right) \mathrm{H}^{\otimes N}.
\end{equation}
Here $C^x_{i,j}$ is the controlled $\sigma_x$-gate with control qubit $i$ and target qubit $j$, and $H$~is the Hadamard gate.

\bibliography{sources}

\end{document}